\documentclass{sf2a-conf2014}
\usepackage{graphicx}
\usepackage{hyperref}
\usepackage[]{natbib}  
\usepackage{epstopdf}

\def\BibTeX{{\rm B\kern-.05em{\sc i\kern-.025em b}\kern-.08em
    T\kern-.1667em\lower.7ex\hbox{E}\kern-.125emX}}
\bibpunct{(}{)}{;}{a}{}{,}  %%%%%%%%%%%%%  A&A bibliography style
\begin{document}

\TitreGlobal{SF2A 2014}

%%-----------------------------------------------------------------
%%      the top matter
%%

\title{Galaxy Formation History Through Hod Model From Euclid Mock Catalogs }

\runningtitle{Galaxy Formation Through HOD}

\author{Z. Sakr$^{1,2,}$}\address{NDU, Louaizeh Zouk Mosbeh, Lebanon}
\address{USJ, Mar Roukoz Dekwaneh, Lebanon}
\address{Laboratoire Lagrange, OCA, CNRS, UNS, boulevard de l'Observatoire 06304 Nice Cedex, France}

\author{C. Benoist$^{3}$}
%% IF Author3 has the same affiliation than Author1:
%\author{C.\,E. Z. Sakr$^2$}

%% IF Author3 has its own affiliation:
%\author{C.\,E. Author3}\address{Dept. of Chess, University of Games, 35101 Las Vegas, Monaco} 

%% IF Author3 has two affiliations, the one of Author1 and a second one:
%\author{C.\,E. Author3$^{1,}$}\address{Dept. of Chess, University of Games, 35101 Las Vegas, Monaco} 

%% Keep this line, even if the page will be settled afterwards.
\setcounter{page}{237}

%%-----------------------------------------------------------------

\maketitle

%%-----------------------------------------------------------------
%%        The abstract
%% 
%%  Warning!  within the abstract:
%%  - do not use macros. 
%%  - do not use commands like: \cite, \citet, \citep ... etc.

\begin{abstract}
Halo Occupation Distribution (HOD) is a model giving the average number of galaxies in a dark matter halo, function of its mass and other intrinsic properties, like distance from halo center, luminosity and redshift of its constituting galaxies. It is believed that these parameters could also be related to the galaxy history of formation. We want to investigate more this relation in order to test and better refine this model. To do that, we extract HOD indicators from EUCLID mock catalogs for different luminosity cuts and for redshifts ranges going from $0.1 < z < 3.0$. We study and interpret the trends of indicators function of these variations and tried to retrace galaxy formation history following the idea that galaxy evolution is the combination rather than the conflict of the two main proposed ideas nowadays: the older hierarchical mass merger driven paradigm and the recent downsizing star formation driven approach.
\end{abstract}

%% Insert the keywords (to appear in the ADS indexing)
%% Keywords must be separated by a comma
\begin{keywords}
Halo Occupation Distribution, EUCLID, Mock catalogs, Galaxy Formation
\end{keywords}

%%-----------------------------------------------------------------

\section{Introduction}
%%---------------------
  Long time passed before advances in the theory of dark matter halo formation (DMH) by hierarchical mass merger
driven process and its relation to galaxy formation (from the fact that inflow of gas into DMH potential well to a high cold gas density
\citep{1978MNRAS.183..341W} could trigger star formation) could be tested through N body simulations combined with
semi analytic approach \citep{1993MNRAS.262..627L}. Many advances in trying to model galaxy halo's number or the Halo Occupation Distribution (HOD) will
follow after but it was mainly  \cite{1999MNRAS.303..188K} and \cite{2000MNRAS.311..793B} who stated first that the average number of galaxies in a
given DMH, which is directly related to the HOD, depends as a power law on its mass. This law has been later refined to explain
why it breaks on small and very large scale by taking into account the role of other parameters, like distance of galaxies from halo center, thus
dividing them into big massive luminous centrals and smaller satellites \citep{2002ApJ...575..587B,2004ApJ...609...35K}, 
or luminosity of halo's constituting galaxies \citep{2005ApJ...633..791Z}. Attempts also where made to
include evolution of halo's number of progenitors through redshift \citep{2007ApJ...667..760Z}.

Several groups
\citep{2005ApJ...630....1Z,2007ApJ...667..760Z,2010MNRAS.406.1306A,2012A&A...542A...5C} have tried to investigate
galaxy formation by studying HOD obtained from a fit to a correlation function extracted from different surveys. We aim at doing
the same with the difference that we compute HOD directly from mock catalogs constructed by \cite{2013MNRAS.429..556M}
from simulations of future observations by  EUCLID space mission. This will be a test of the upcoming EUCLID mission and an attempt to extent works cited before as none of them have used a sample of galaxies as large and deep at the same time as EUCLID, with redshift reaching $z \sim 3.0$ and potential galaxy number observed, in the order
of $50\times10^6$ (Euclid Definition Study Report \citeyear{2011arXiv1110.3193L}).

Many concordant evidences and observations \citep[see][and references therein]{2012RAA....12..917S}, have
helped establish a hierarchical theory of galaxy formation as a continuation to the DMH bottom up scenario of large scale structures
evolution. This theory (from \cite{1978MNRAS.183..341W}; \cite{1991ApJ...379...52W}to \cite{2006ApJ...652..864H,2008ApJS..175..356H}) has been challenged by
other observational data of galaxy mass downsizing from $z \sim 1-2$ zone down to low redshifts \citep{1996AJ....112..839C}. This led \citet{2004Natur.428..625H}
and \citet{2006MNRAS.366..499D} to suggest that it is due to the fact that most early type massive galaxies 
stop forming stars first due to different quenching processes, while late type lower mass ones remains
active and become quiescent later \citep{2003MNRAS.341...33K}.

In this study, we extract HOD's mean galaxy number for different luminosity cuts and redshifts ranges. We then calculate
for each extraction its specific indicators like $M_{min}$ (resp. $M_{amp}$) mass of halo hosting one central (resp.
satellite) galaxies, the index $\alpha$ of the power law, the average halo mass $\bar{M}_{
halo}$ weighted by galaxy number and galaxy satellites fraction $f_{sat}$. After that, we try to interpret the change in their trend, function of redshift and luminosity, in the
light of the previously advanced ideas of galaxy formation.
  
\section{Data selection}
%%-------------------------

EUCLID is a space telescope developed by ESA to be launched in 2019. It will perform visible and near-infrared
imaging up to $24.5 \ mag$ apparent magnitude and NIR spectroscopy in $AB$ system for wavelength range going
from 460 $nm$ to 2000 $nm$. This will allow him to scan $\sim 50\times10^6$ galaxies in a large region of 15,000 $deg^2$ with
depth reaching $z\sim3$ (see Euclid Definition Study Report \citeyear{2011arXiv1110.3193L}). To test the benefit of such an unprecedented
deep and large survey on galaxy history of formation through cosmic time, we used mock catalogs constructed by \citet{2013MNRAS.429..556M}.
These mock catalogs were constructed by grafting a semi-analytic
model of galaxy formation, GALFORM from \citet{2011MNRAS.418.1649L} onto the N-body dark matter halo merger trees
of the Millennium Simulation by \citet{2005Natur.435..629S}. From the different outputs of these constructions we use
the EUCLID 100 Hband DEEP lightcone implemented using the Lagos12 GALFORM model. The lightcone
covers the redshift range $z \sim 0.0$ to $z \sim 3.0$ and has a sky coverage of 100.21 $deg^2$, with an apparent magnitude
cut $m < 27 \ mag$ and a cosmology of $\Omega_m = 0.25$; $\Omega_{\Lambda} = 0.75$; $h = 0.73$; $n_s = 1$; $\sigma_8 = 0.9$. We want to extract
the HOD from our mock catalog to study how this distribution vary according to halo mass of course, but also
redshift range and luminosity cut. We take redshift bin to be $\Delta z \sim 0.1$. This range will allow us first to
spot changes in trends related to galaxy formation and evolution from local universe to redshift $z \sim 1$ as well as when passing to $z \sim 1-2$ zone and higher. We move
next to the luminosity criteria and begin with an absolute $H$ band magnitude range between $-20 > M_H > -21$ for
all redshift limited samples. We stay on a stable number of galaxies within this magnitude variation which is also
above the threshold brightness that insure completeness for all the samples in our redshift ranges. Taking
these considerations into account, we varied this luminosity range by $\Delta M \sim 0.1$ to get more samples and compare their plots
of variation. We come at the end to the choice of the mass bin. The whole mass range up to $ \sim 10^{15} M_{\odot}$ will
be divided to 500 bins. This is small enough to detect the HOD indicators mentioned before, which are in
the order of $M_{min} \sim 10^{11} M_{\odot}$ and $M_{amp}  \sim 10^{13} M_{\odot}$ and large enough to insure the robustness of the bin as a
sample of number of halos. We also limit ourselves to $10^{14} M_{\odot}$ as upper limit as the number of halos above
that value drops below 10 (Left Panel of Fig.~\ref{sakr:fig1}) and the systematic statistical error becomes higher than
10\%

\section{Method and results}
%%-------------------------

\begin{figure}[ht!]
 \centering
 \includegraphics[width=0.48\textwidth,clip]{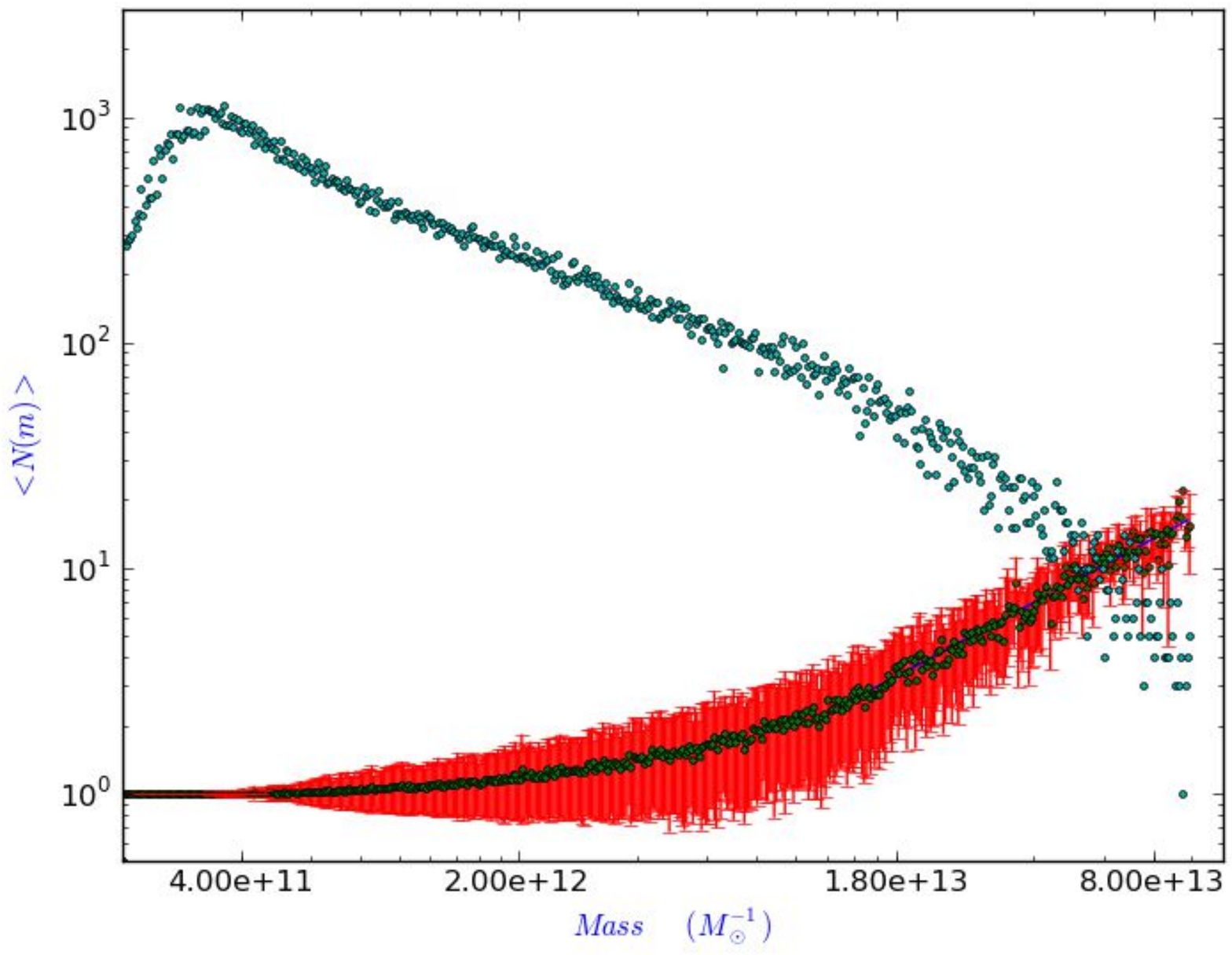}%      
 \includegraphics[width=0.48\textwidth,clip]{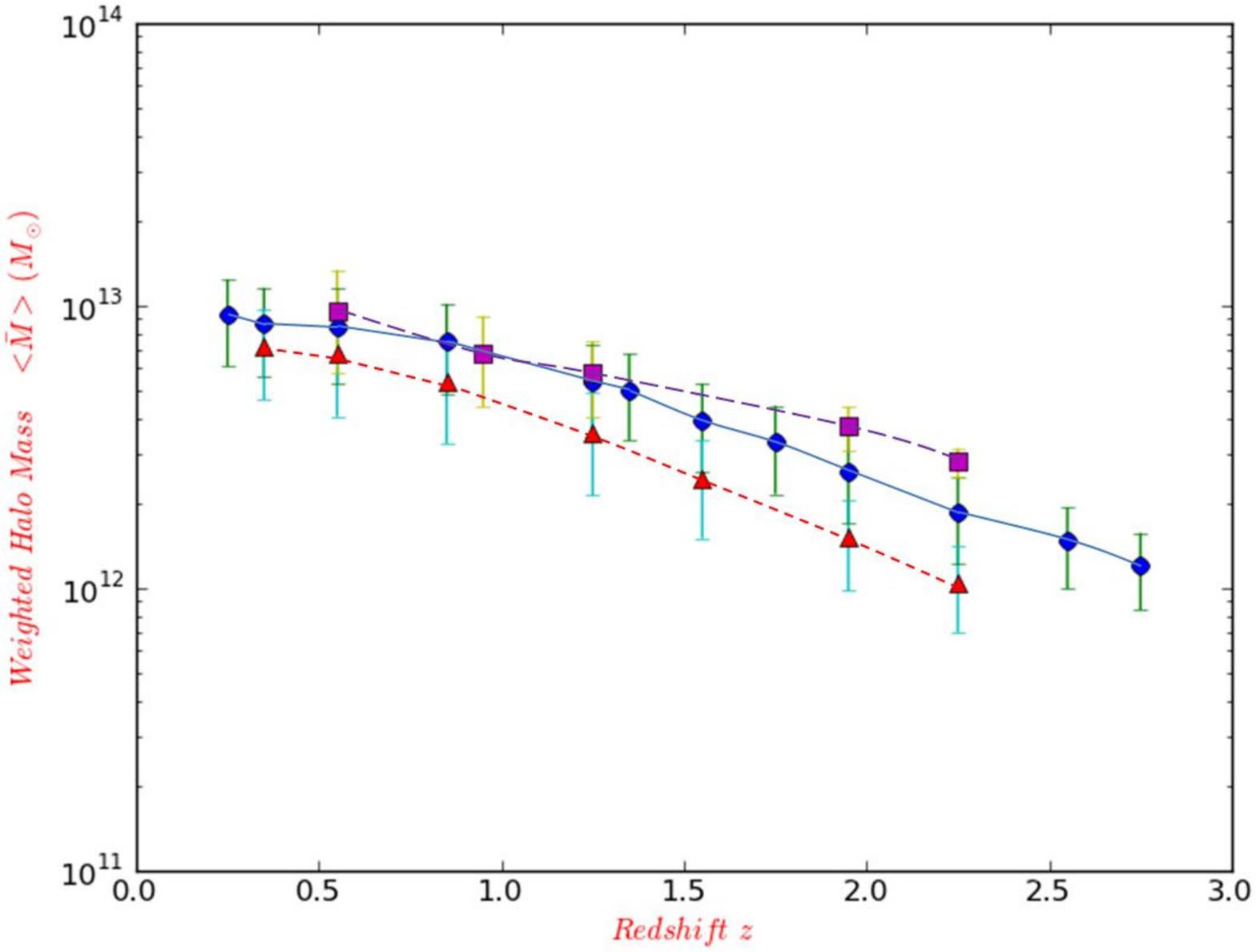}      
%% Note the ABSENCE of the extension .pdf  !
  \caption{{\bf Left:} mean galaxy number per halo mass plot (Red Dots) with halo count (Blue Dots) for $0.8 < z < 0.9$ having $-20 > M_H > -21$. {\bf Right:} weighted halo mass function of redhsift with $-18.5 > M_H > -19$ (Dotted Line) $-20 > M_H > -21$ (Solid Line) $-21.5 > M_H > -22$ (Dashed Line) }
  \label{sakr:fig1}
\end{figure}

\begin{figure}[ht!]
 \centering
 \includegraphics[width=0.48\textwidth,clip]{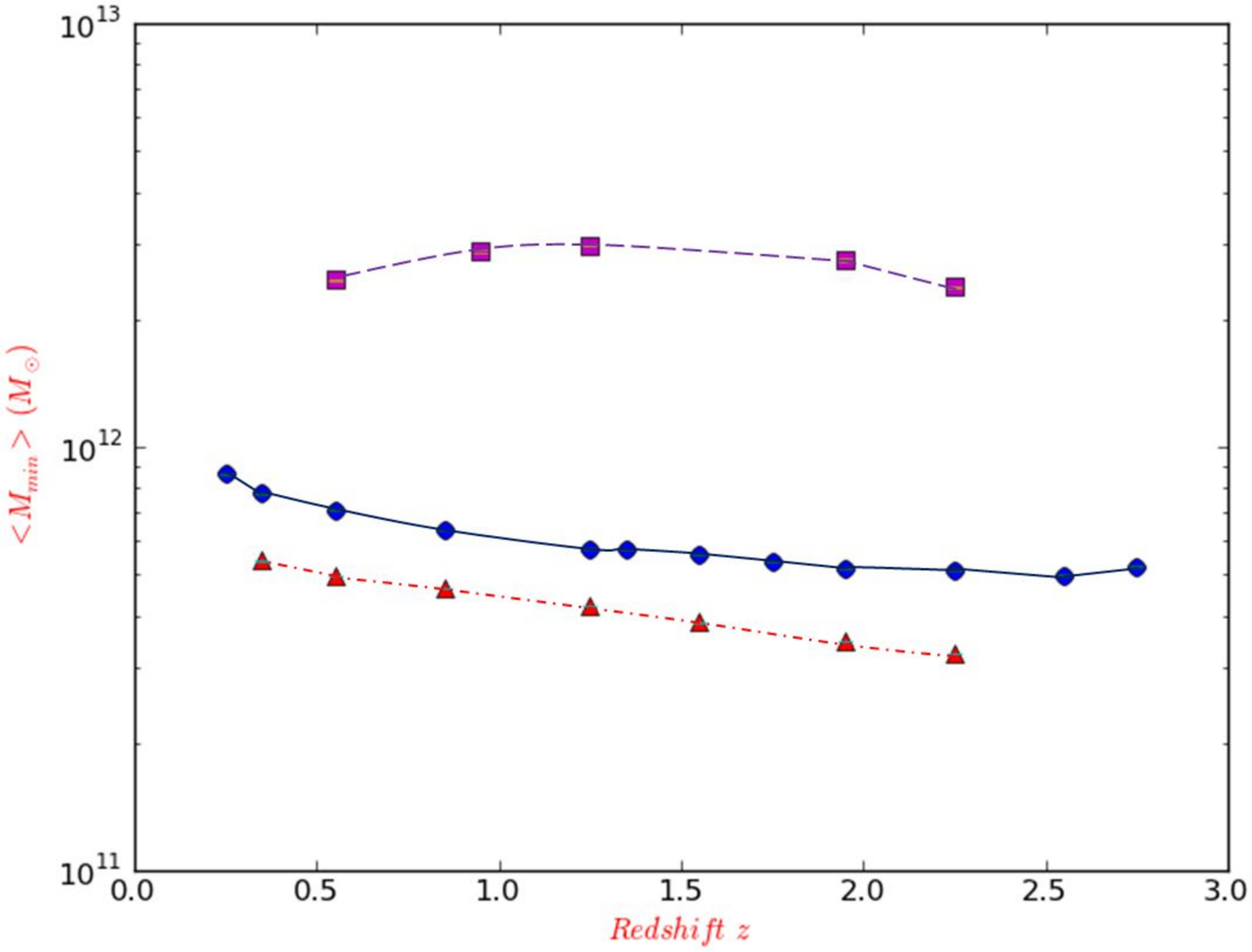}%      
 \includegraphics[width=0.48\textwidth,clip]{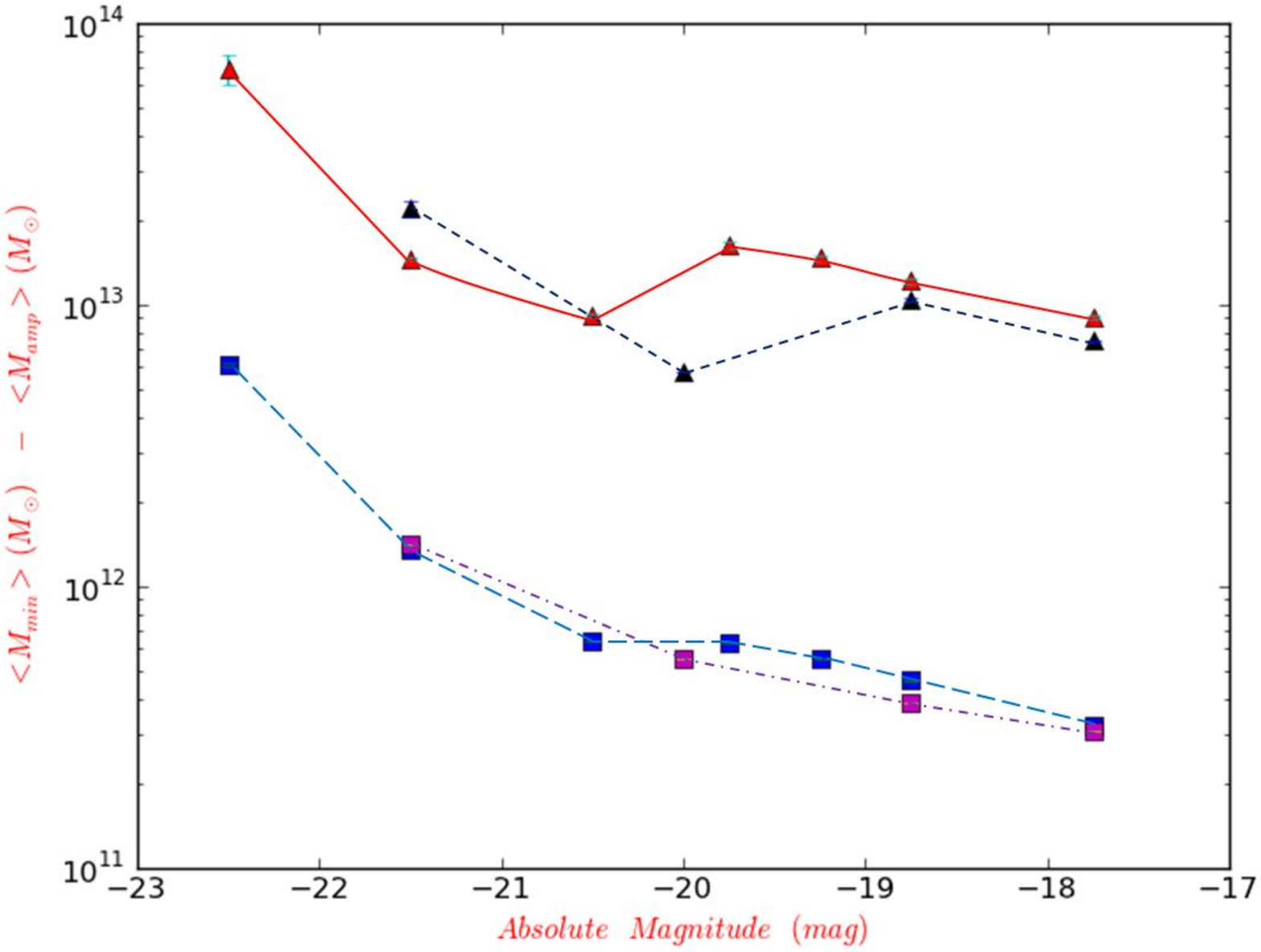}  
 \includegraphics[width=0.48\textwidth,clip]{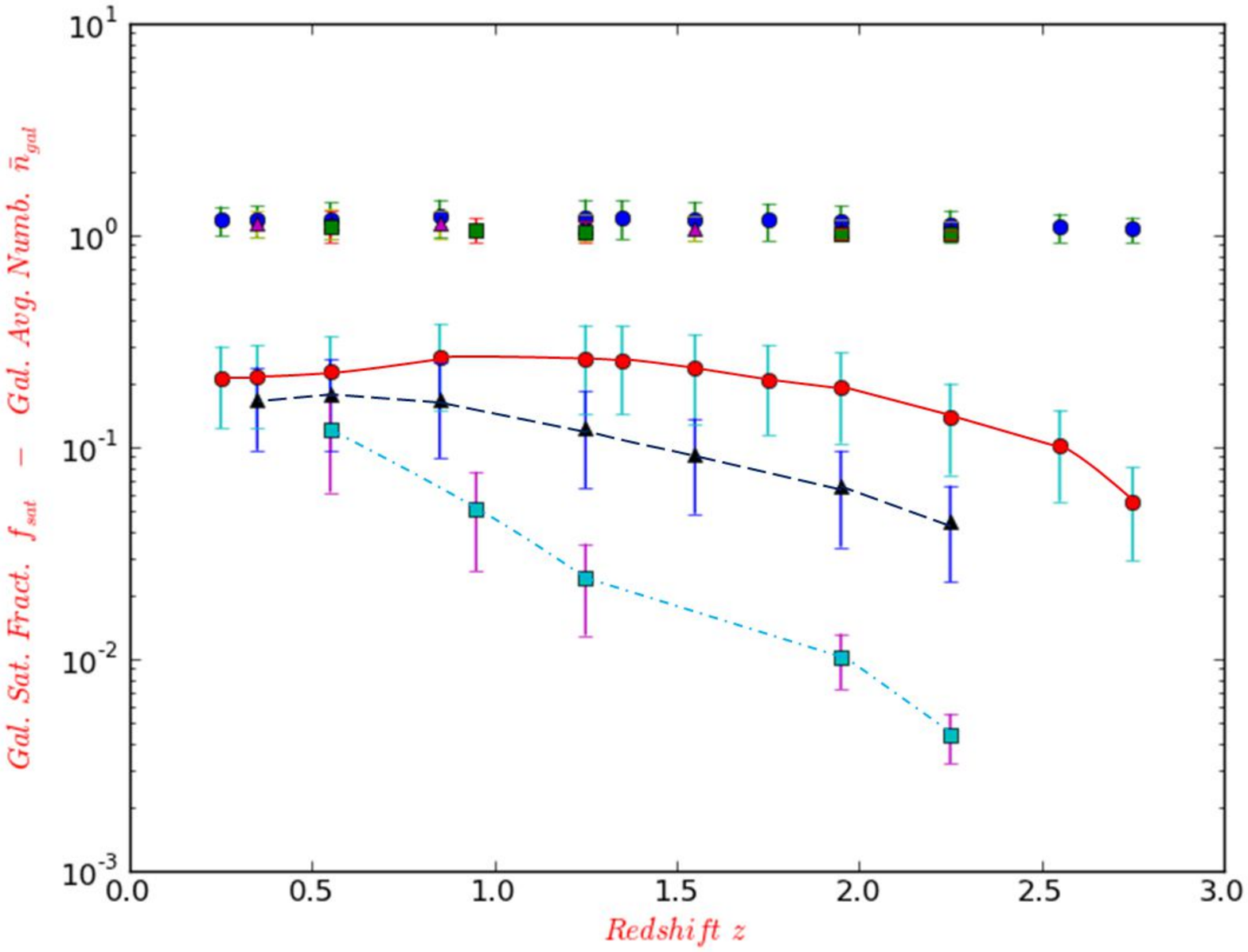}%      
 \includegraphics[width=0.48\textwidth,clip]{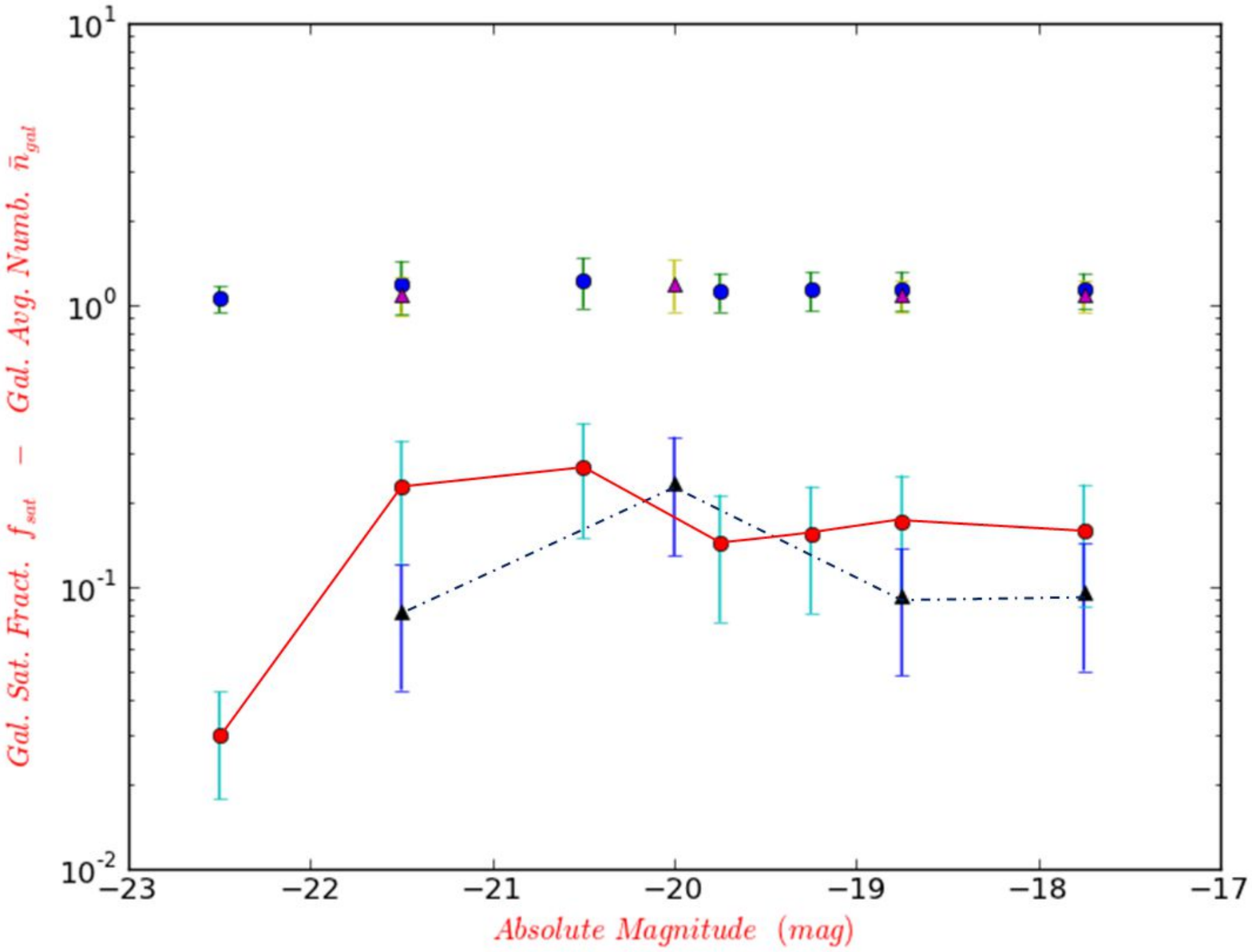}     
%% Note the ABSENCE of the extension .pdf  !
  \caption{{\bf Top Left:} $M_{min}$ function of z for different luminosity cuts. $-18.5 > M_H > -19$ (Dotted Line) $-20 > M_H > -21$
(Solid Line) $-21.5 > M_H > -22$ (Dashed Line). {\bf Top Right:} $M_{min}$ and $M_{amp}$ function of luminosity for different $z$.
$0.8 > z > 0.9$ (Solid Line for $M_{amp}$ dashed for $M_{min}$ ) $1.5 < z < 1.6$ (Dotted Line for $M_{amp}$ dash-dotted for $M_{min}$ ).{\bf Down Left:} galaxy fraction and satellite fraction function of $z$ for different luminosity cuts: $-18.5 > M_H > -19$ (Dotted Line) $-20 > M_H > -21$ (Solid Line) $-21.5 > M_H > -22$ (Dashed Line). {\bf Down Right:}galaxy fraction and satellite fraction function of luminosity for different $z$: $0.8 < z < 0.9$ (Solid Line) $1.5 < z < 1.6$ (Dashed Line) }
  \label{sakr:fig2}
\end{figure}

To model HOD, we use \citet{2002ApJ...575..587B} and \citet{2004ApJ...609...35K} parametrization $\left\langle N(M) \right\rangle = 1$
for $M > M_{min}$, the minimum halo mass for hosting one central galaxy and $\left\langle N(M) \right\rangle = 1 + (M/M_{amp})^{\alpha}$ for
$M > M_{crit}$, $M_{amp}$ being the mass above which the halo could host a satellite. After calculating mean galaxy number per halo mass for
samples chosen according to the previous section (see Left of Fig.~\ref{sakr:fig1} as example for one redshift range), we calculate Mmin, Mamp and $\alpha$, then extract
three more indicators : weighted average halo mass $\bar{M}_{halo}$, galaxy average number per halo $\bar{n}$ and galaxy satellite
fraction $f_{sat}$ and represent their variation in function of $z$ (Left Panel of Fig.~\ref{sakr:fig2}) or in function of luminosity
(Right Panel of Fig~\ref{sakr:fig2}).\\

To summarize, we say that $M_{min}$ and $M_{amp}$ decrease from high $z$ to touch a bottom at $ z \sim 1.5-2$
before rising a little again after, with a linear correlation between $M_{min}$ and $M_{amp}$ $\sim 15-18$ in accordance with
\citet{2005ApJ...633..791Z} simulations studies and \citet{2012A&A...542A...5C} observations studies for lower values of redshift.
These trends are consistent with those found in both
the local \citep{2005ApJ...630....1Z,2011ApJ...736...59Z} and distant Universe studies \citep{2007ApJ...667..760Z,2010MNRAS.406.1306A}. 
Also they concord in the general trends with results on observations
between $ z \sim 0.2$ and $ z \sim 1.2$ done by \citet{2012A&A...542A...5C}. \\

As a first general interpretation (more thorough analyzes in upcoming Sakr \& Benoist paper) of these trends we say that combining the hierarchical and the
downsizing theory could account for most of their behavior. We divide the redshift range in three parts,
$2.0 < z < 3.0$, in which galaxy increase formation rate and increase mass is fueled by high merger rate of early
structure formed in high density peaks along with active star formation of the still young galaxies, $2.0 < z < 1.0$,
where this process culminate and stabilize with downsizing effect beginning to show and finally $1.0 < z < 0.0$
where big merger rate and new born galaxies drops and early galaxies type quench star formation while late type
small are still active resulting in a downturn of the previous trend (not the absence of this behavior for  high luminosity cuts leaving only massive early type galaxies that follow the hierarchical theory). This conciliates discrepancies mentioned previously
and concords with the same trend observed for the three zone for star formation rate \citep{2012A&A...539A..31C} or
galaxy pair merger rate established by \citet{2008MNRAS.386..909C} with a pivot at $z \sim 1$. It accounts also for the decrease of the rate
of big mergers noticed by \citet{2009A&A...498..379D} along with an increase of minor mergers from \citet{2010ApJ...710.1170L}. 
It is also consistent with the halo mass distribution function of redshift \citep{2004ApJ...609...35K} suggesting 
an increase with low $z$ in the number of small size DMH 'incubation' containers resulting in low mass galaxies forming in a rate higher than for the massive ones.

\section{Conclusions}
%%--------------------

The results obtained, showed that we can extend HOD model from only a manifestation of the hierarchical
theory of galaxy formation to include other suggested ideas like, as we tried to do, the newly supported by many
observational evidences, downsizing approach. However this couldn't be done without calculating the variation of
HOD's related indicators over a large range of redshift and luminosity. This show the need of conducting large
deep spectroscopic surveys like the future Euclid space mission where no restrictions coming from the need 
to maintain a specific criteria can filter the large
population observed to insignificant statistical samples. Also these results could serve as a test for an eventual scientifically meaningful 
model that will parametrize HOD according to redshift, as such an operation could give more precise physical
meaning to the trends we obtained and help clarify many issues related to galaxy formation.

%% The following lines are required when using BibTEX (strongly encouraged!):
\bibliographystyle{aa}  % A&A bibliography style file (aa.bst)
\bibliography{sakr} % your references in file: Yourfile.bib

\end{document}